# Crystal Growth, Band Structure, Magnetism and Electrochemical Properties of Hexavalent Strontium Ruthenium Oxyhydroxide

*Subham Naik,[a] Soumili Dutta,[a] Hiranmayee Senapati,[a] Sweta Yadav,[c] Subarna Ray,[b] Jai Prakash,[c] Rahul Sharma,[b] Gohil S. Thakur[a*]*

[a]Department of Chemical Sciences, Indian Institute of Science Education and Research, Berhampur-760030, Odisha, India

[b]Department of Physical Sciences, Indian Institute of Science Education and Research, Berhampur-760030, Odisha, India

[c]Department of Chemistry, Indian Institute of Technology Hyderabad, Kandi, Sangareddy-502284, Telangana, INDIA

**Abstract**

Ruthenates comprise an interesting class of materials with a wide range of extremely exciting properties, and thus the discovery of new stable ruthenates remains an active area of investigation. We report the crystal growth and comprehensive studies including crystal and electronic structure, magnetic and electrochemical properties of a hexavalent ruthenium oxyhydroxide $Sr_3Ru_2O_8(OH)_2$ prepared through a low-temperature hydrothermal method. Single crystals and powder samples of this phase are isolated by optimising the $Sr(OH)_2$ to $KRuO_4$ ratio while maintaining a high base concentration. The new structure consists of a rare five-coordinated $Ru^{VI}$ featuring isolated trigonal prisms and crystallising in a non-centrosymmetric tetragonal system. Isolated Ru polyhedra leading to a large spatial distance (~ 5 Å) between the Ru metal centres render the compound paramagnetic despite strong antiferromagnetic correlation. Band structure calculation suggests a metal-like electronic ground state with mostly Ru *d* and O *p* orbitals contributing to the Fermi surface. The electrochemical performance of $Sr_3Ru_2O_8(OH)_2$, though not as impressive as $RuO_2$, remains relevant and is on par with other reported OER catalysts.

**Introduction**

Ruthenates are recognized as prominent materials within the realms of condensed matter physics and materials science, offering significant opportunities for the investigation of emergent phenomena arising from the complex interactions of charge, spin, orbital, and lattice degrees of freedom. In this context, the perovskite family and its derivatives hold particular significance due to their remarkable diversity in electronic,[1] magnetic,[2] and structural properties. These attributes can be precisely manipulated through careful modifications in composition and dimensional parameters.[3] Among this expansive landscape, strontium ruthenates, which constitutes the layered Ruddlesden-Popper (RP) homologous series characterized by the general formula $Sr_{n+1}Ru_nO_{3n+1}$, have emerged as critical system for both fundamental research and potential technological advancements. This series provides a distinct platform for examining the effects of dimensionality and strong electron correlations on material properties,[3] progressing from the three-dimensional perovskite $SrRuO_3$ (n = ∞)[4] to the quasi-two-dimensional single-layer $Sr_2RuO_4$ (n = 1)[5] and bilayer $Sr_3Ru_2O_7$ (n = 2)[5] structures.

Conventional solid-state synthesis methods provide access to only a limited thermodynamically stable phases, and competing phases often result in a sample with multiple phases. However, low-temperature methods like hydrothermal synthesis provide an alternate platform to discover new kinetically stable phases (often single-crystalline) that are otherwise inaccessible through high-temperature solid-state synthesis. A significant number of ruthenates have been reportedly grown under mild hydrothermal conditions in the last 15 years.[6–10] Frequently, phases with metal in higher (+4, +5) or mixed oxidation states have been achieved. However, purely hexavalent ruthenates are very rarely encountered.

Most of the reported ruthenates contain octahedra in various connectivity, giving rise to large structural diversity, whereas, among the known hexavalent ruthenates, two particular structural classes emerge: 1) alkali ruthenates ($A_2RuO_4$, A = K, Rb, Cs)[11,12] exhibiting *β*-$K_2SO_4$ type structure, or $A_3A'(RuO_4)_2$ (A = Rb, K and A` = Na)[13] featuring a $RuO_4$ tetrahedra [with $Na_2RuO_4$[14] and related $Ag_2RuO_4$[8] as exceptions, showing chains of vertex-connected $RuO_5$ trigonal pyramids (TBP)] and 2) $A[RuO_3(OH)_2]$ (A = Ba, Sr, $K_2$),[15,16] including $BaHgRuO_5$,[17] possessing isolated $RuO_5$ trigonal pyramids. $CsK_5(RuO_4)(RuO_5)$[18] contains both $RuO_4$ tetrahedra and $RuO_5$ TBP and $Ag_2RuO_4$,[19] despite a composition similar to $K_2RuO_4$, a TBP geometry is adopted. It appears that in Ru(VI) oxides, both the coordination (tetrahedra and TBP) occurs with equal probability, and an octahedral coordination is categorically avoided. Until now, all these higher-valent (higher than +5) oxo(hydroxo)ruthenates listed above have been prepared under strong oxidising conditions (use of strong oxidiser, high oxygen pressure or oxygen annealing at high temperatures). To the best of our knowledge, isolation of a purely hexavalent ruthenate by low temperature or '*chimie douce*' method has been reported only for $Ag_2RuO_4$.[19] Here we present crystal growth, structural, magnetic and electrochemical properties of a hexavalent hydroxy ruthenate $Sr_3Ru_2O_8(OH)_2$, crystallized under low-temperature hydrothermal conditions by varying the initial metal ratio (Sr/Ru) at a constantly high pH. It crystallizes in a new structure featuring isolated trigonal pyramids of ruthenium. This study highlights the importance of exploratory hydrothermal synthesis as a powerful tool to isolate novel phases with uncommon oxidation states and coordination preferences.

**Materials and Methods**

*Crystal Growth*: Single-crystalline and phase-pure powder samples of the material were grown via a hydrothermal method. An optimized molar ratio of 5:2:4 was maintained for the precursors: strontium hydroxide octahydrate [$Sr(OH)_2$, Sigma Aldrich 99.9 %], potassium perruthenate ($KRuO_4$, Sigma, Ru ≥ 49%), and sodium hydroxide (NaOH, SRL, 97%). These compounds were accurately weighed and mixed in a 25 ml Teflon vessel with 15 ml deionized water. The mixture was stirred for 30 minutes, yielding a brown-coloured solution with a pH of ~13. The solution mixture was hydrothermally treated at 180°C for 72 hours.

Post-reaction, an orange solution (pH ~ 13) accompanied by a black precipitate was observed. The precipitate was isolated by filtration, followed by multiple washings with distilled water until a neutral pH was attained to ensure all the NaOH had been washed away. Finally, the sample was rinsed with methanol and dried under vacuum overnight. The final product comprised of polycrystalline powder and many cuboid-shaped crystals, which were manually separated for subsequent structural and compositional analysis.

The morphology and elemental composition of the grown crystals were analysed using a field emission scanning electron microscope (FESEM, model and make: JEOL, Japan, JSM 7610F) equipped with an octane elite energy dispersive X-ray (EDX) spectrometer. Data were collected from multiple points on several crystals, and the atomic percentages were averaged to determine the semiquantitative stoichiometric ratio of the elements. No sodium was detected in any of the crystals.

*Single Crystal X-ray Diffraction*: Single crystal diffraction data were collected at 298 K using a Bruker diffractometer (model: D8 Venture) equipped with a Mo-Kα (λ=0.71073 Å) source and a Photon III mixed-mode X-ray detector. The APEX-3 software was used for data acquisition and data extraction.[20] Data reduction and cell refinement were done using Bruker SAINT software. The SHELXT 2014/5 program is used to solve the structure using the intrinsic phasing method, which is based on the direct and dual space methods,[21] and the SHELXL 2019/1 program is used to refine the final structure using Least Squares minimisation.[22,23] The multi-scan method of the SADABS program was used for the absorption corrections.[24] Detailed crystallographic data and structural refinement parameters are summarised in Tables S2-S4. The crystallographic data have been submitted to the joint CCDC/FIZ Karlsruhe deposition service with the CSD number 2465126. The CIF file may be obtained free of charge by contacting the CCDC at https://www.ccdc.cam.ac.uk/.

*Powder X-ray Diffraction:* The powder diffraction data were collected using a Malvern Panalytical X-ray diffractometer (Empyrean 3), equipped with a PIXcel 3D area detector and using copper Kα (1.54 Å) X-ray source. Data were collected through three scans over a 2θ range of 5 to 90°, at a step size of 0.013° and an approximate duration of 75 seconds per step.

*X-ray Photoelectron Spectroscopy (XPS):* The photoelectron spectra (XPS) of the sample were obtained using ULVAC-PHI Inc., Phi Genesis X-ray photoelectron spectrometer with a monochromatic X-ray source Al Kα (1486.6 eV; 25 W, 15 kV). The peaks were calibrated using the C 1s peak as a standard with a binding energy value of 284.5 eV.[25] The survey scan was conducted within a binding energy range of 0 to 1100 eV. The sample was kept in the high-vacuum chamber of the instrument, approximately $1.6 \times 10^{-6}$ Pa, overnight. Data collection consisted of 10 cycles, with each cycle featuring 12 sweeps at a duration of 50 ms per sweep. Sr 3d, Ru 3d, and O 1s data are collected for our titled compound.

*Fourier-transformed Infra-red Spectrum (FTIR):* The transmittance spectra of the crystals were acquired utilizing a Shimadzu IRSpirit-ZX spectrometer, which is equipped with a QATR-S single reflection ATR accessory. The data were obtained in transmission mode across a 500 to 4700 cm$^{-1}$ spectral range. A total of 45 scans were performed, achieving a resolution of 8 cm$^{-1}$. The background correction is done by taking the spectra without any sample and subtracting them from the spectra of the crystals.

*Thermal Analysis:* The thermogravimetric data for the sample were collected using Stare software on a Mettler Toledo TGA2 thermogravimetric analyzer, which includes a small furnace, an XP1 balance, and a TGA sensor. The data were obtained over a temperature range of 303 to 1350 K in both nitrogen (N$_2$) and ambient environments, at a heating rate of 10 K/min. Background correction was performed using data from a blank 70 μL alumina crucible.

*Magnetic Property Measurements:* The DC magnetization was measured as a function of temperature (T) and magnetic field (H) using a Vibrating Sample Magnetometer (VSM) on a Cryogen Free Measurement System (CFMS, Cryogenics). Susceptibility data were collected at temperatures ranging from 2 K to 300 K, in an applied magnetic field of 0.5 T in both zero-field-cooled and field-cooled (ZFC and FC) protocols. AC- susceptibility data in the limited t-range of 5-35 K at various applied frequencies were recorded on an MPMS SQUID.

*Electrochemical Measurements:* Electrode preparation and electrochemical testing were conducted using a CHI 760E electrochemical workstation. The electrodes consisted of reference (Ag/AgCl), counter (platinum), and working (catalyst-coated graphite sheet) electrodes arranged in a three-electrode configuration. The catalyst was applied to the graphite sheet electrode at a loading of 0.5 mg/cm² using an ink composed of 500 μL of Milli-Q water, 5 mg of catalyst, and 50 μL of Nafion perfluorinated resin as a binding agent. The coated electrodes were dried under vacuum. electrochemical measurements were performed via linear sweep voltammetry (LSV) in a 1 M KOH electrolyte, with voltages ranging from 0 to 1 V at a scan rate of 10 mV/sec. For evaluating water oxidation activity, LSV was performed at a scan rate of 10 mV/sec in the potential range of 1-2 V vs. RHE. The equation used in the conversion of electrochemical data into RHE: $E_{RHE} = E_{Ag/AgCl} + 0.197 + 0.059*pH$

*Band Structure calculation:* DFT calculations were conducted using the Quantum ESPRESSO software,[26,27] employing ultrasoft pseudopotentials to model valence electron-ion interactions. The valence configurations used for the elements were as follows: Sr ($4s^2\ 4p^6\ 5s^2$), Ru ($4s^2\ 4p^6\ 4d^6\ 5s^2$), O ($2s^2\ 2p^4$), and H ($1s^1$). A kinetic energy cutoff of 70 Ry and a charge density cutoff of 420 Ry were applied to expand the wave function and charge density on a plane wave basis.[28] The Perdew-Burke-Ernzerhof (PBE) parameterization was utilized to describe the electron-electron exchange and correlation interactions.[29] A $6 \times 6 \times 12$ Monkhorst-Pack *k*-point mesh was implemented for Brillouin Zone integrations during self-consistent calculations.[30] For non-self-consistent calculations, a denser *k*-point grid of 14 × 14 × 28 was used, with an energy convergence threshold of 1 × 10$^{-9}$ Ry. The structure was optimized and visualized using VESTA software.[31]

**Results and Discussion**

*Crystal morphology and composition:* Black cuboid crystals and polycrystalline powder were formed under hydrothermal conditions. The obtained crystals have size up to 90 μm (in the longest dimension), as estimated from optical images and FESEM micrographs (Figure 1). Elemental analysis yielded a Sr:Ru ratio of approximately 3:2, which matches the composition determined from single-crystal x-ray data. Elemental ratio is presented in Table S1. Sodium was not detected in any of the crystals.

*Phase formation:* Extrapolating the speciation diagram presented by Tarascon et al,[7] it is inferred that on increasing both the Sr/Ru ratio and the [OH] concentration, a composition with ruthenium valency higher than 5+ may result, and that to synthesize alkaline-earth rich ruthenate, a higher [OH] concentration with a higher Sr/Ru ratio is required. This motivated us to vary the metal ratio at a high [OH] concentration (pH ~ 13).[6] Various starting metal ratios, Sr/Ru (0.5 - 2.5), were tried while keeping the quantity of NaOH constant. At a starting metal ratio of 1, the primary phase isolated was $Sr_3[RuO_4(OH)]_2$ (blackish/brown block-shaped crystals) along with significant amounts of $SrRuO_4·H_2O$ as the side phase (dark orange hexagonal bipyramidal crystals). Adjusting the Sr/Ru ratio to 2.5 resulted in a phase-pure $Sr_3[RuO_4(OH)]_2$ sample. A comparison plot of the powder diffractogram of our sample with the calculated powder diffraction pattern from SCXRD is presented in Figure 1.

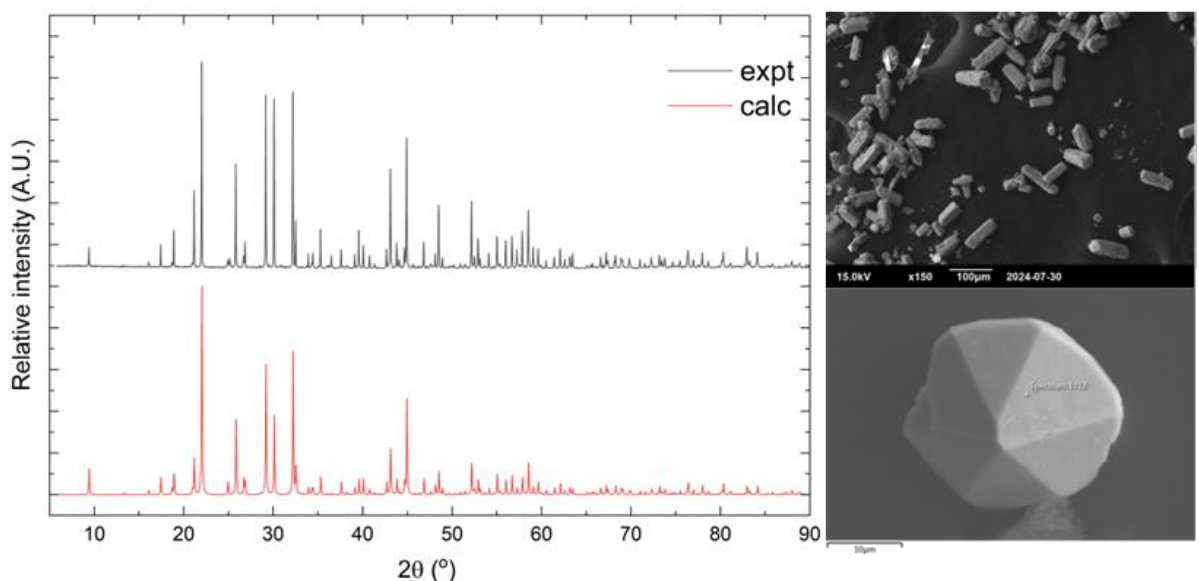

**Figure 1.** Experimental powder X-ray pattern of $Sr_3[RuO_4(OH)]_2$ compared to the pattern simulated using single crystal data and SEM images of the obtained crystals.

Sr/Ru ratio ≤ 1, resulting in exclusively $SrRu_2O_6$ phase in accordance with previous reports.[6] In some synthesis batches with a smaller Sr/Ru ratio, Hexagonal pyramidal crystals with a metal ratio of 1:1 were observed, presumably belonging to a reported but not properly studied phase $SrRuO_4·H_2O$, as suggested by the few extra reflections observed in the PXRD (PDF no. 35-949).[32] However, due to small size, poor crystal quality compounded by heavy twinning, the structure could not be solved using single crystal data. Attempts to obtain phase pure sample were not made.

*Crystal Structure.* The crystal structure was solved in a primitive tetragonal cell using single-crystal X-ray diffraction data. Crystal data, atomic positions, anisotropic thermal parameters and important bond lengths are listed in Tables 1, and Tables S2, S3 and S4 in the SI. The compound crystallises in a tetragonal cell in a non-centrosymmetric space group $P\bar{4}$ with lattice parameters a = 1324.70(6) and c = 549.72(4) pm.

**Table 1**. Crystal data of $Sr_3[RuO_4(OH)]_2$.

*Crystal data*

| | |
|---|---|
| Wavelength, λ (pm) | Mo Kα, 071.073 |
| Temperature (K) | 298(2) |
| Crystal dimension (mm) | 0.08 × 0.06 × 0.05 |
| Crystal shape & colour | Block, black |
| Empirical formula | $Sr_3Ru_2O_{10}H_2$ |
| Formula weight (g/mol) | 627.02 |
| Crystal system | Tetragonal |
| Space group (No.), Z | $P\bar{4}$ (#81), 4 |
| Unit cell dimensions (pm) | $a = 1324.70(6)$; $c = 549.72(4)$ |
| Volume (pm$^3$) | $9.646 \times 10^8$ |
| Density (g cc$^{-3}$) | 4.317 |
| Absorption correction | Multi-scan |
| Absorption coefficient (mm$^{-1}$) | 19.57 |
| Transmission | $T_{min} = 0.479$, $T_{max} = 0.746$ |
| $F(000)$ | 1128 |
| 2θ range (°) | 2.2–30.5 |
| Index ranges | $-18 \leq h \leq 15$, $-18 \leq k \leq 18$, $-7 \leq l \leq 7$ |
| Reflections collected | 31353 |
| Independent reflections | 2958 |
| Reflections with I > 2σ(I) | 2713 |
| Parameters, restraints | 138, 0 |
| Absolute structure parameter | 0.034(8) |
| Extinction coefficient | 0.00101(13) |
| Goodness-of-Fit on F$^2$ | 1.073 |
| R and wR values [$F^2 > 2\sigma(F^2)$] | 0.0311, 0.0533 |
| R and wR values for all data | 0.0251, 0.0515 |
| Largest difference peak and hole (e Å$^{-3}$) | 0.74, −0.80 |

Note: Refinement method: Full-matrix least-squares on F$^2$.
Extinction correction: SHELXL-2017/1, $F_c^* = kF_c[1 + 0.001 \times F_c^2\lambda^3/\sin(2\theta)]^{-1/4}$ [22,23]
Absolute structure: Refined as an inversion twin.
$w = 1/[\sigma^2(F_o^2) + (0.0107P)^2 + 1.2927P]$ where $P = (F_o^2 + 2F_c^2)/3$

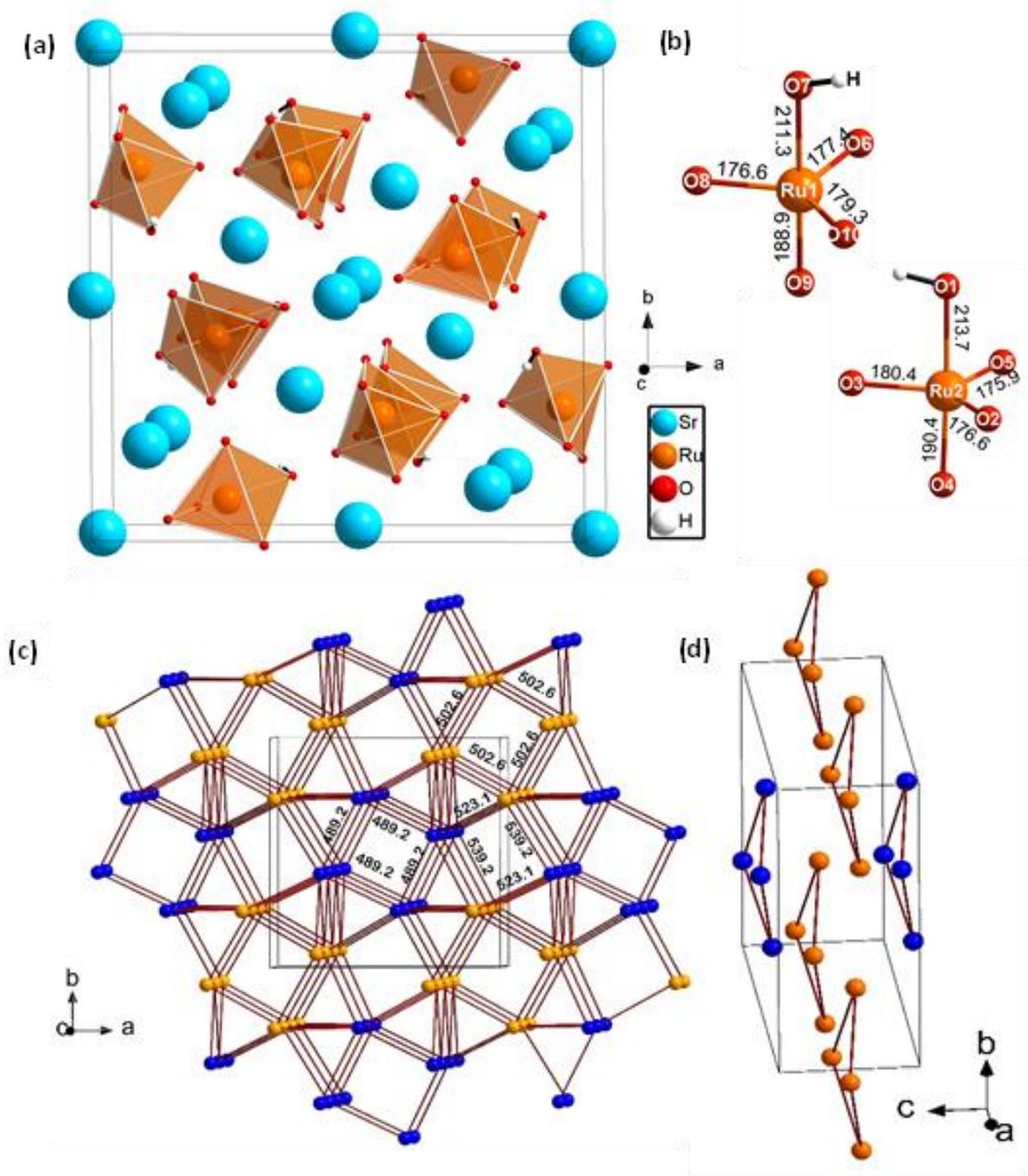

**Figure 2.** Crystal structure of $Sr_3Ru_2O_8(OH)_2$ (a) representative unit cell with Ru polyhedra, (b) Ru-O coordination environment with respective bond lengths (in Å), (c and d) distorted square lattice formed by the two distinct crystallographic Ru atoms.

The structure is composed of isolated $RuO_5$ trigonal pyramids as shown in Figure 2. There are two distinct crystallographic ruthenium atoms; each polyhedra has a different orientation in space. Ru polyhedra have three equatorial bonds in the range of 175-190 pm, which are in agreement with the reported Ru(VI)–O bond distances[12,18,19, 33-35]. While one of the axial bonds is ~190 pm, the other is considerably longer (~211 pm) as presented in Table S4. This suggests there is a possibility of a hydrogen atom attached to the distantly bonded apical oxygen, comprising an –OH group. Therefore, a hydrogen atom is manually attached to each of the O1 and O7 atoms (O–H distance = ca. 0.97 Å) in the structure. Similar anisotropy in bond lengths is observed in $Ba[RuO_3(OH)_2]$[33] and $K_2[RuO_3(OH)_2]$,[16] which contain isolated $RuO_3(OH)_2^{2-}$ ions with –OH groups at both axial positions at a distance of ~ 211 pm. While the octahedral coordination is ubiquitous for $Ru^{4+}$ and $Ru^{5+}$, only a few compositions, $Ba[RuO_3(OH)_2]$,[33] $BaHgRuO_5$,[34] $CsK_5(RuO_3)(RuO_4)$,[18] $Na_2RuO_4$,[35] $K_2RuO_4$,[12] and $Ag_2RuO_4$[19] are known for hexa-valent ruthenates in TBP coordination. The calculated bond-valence sum of 5.85 for Ru1 and 5.78 for Ru2, calculated using the $r_o$ values for $Ru^{6+}$ used for $Ag_2RuO_4$,[19] strongly supports an oxidation state of +6 for both the Ru ions. The two ruthenium atoms (blue and yellow in Figure 2c and 2d) sit on a highly distorted square lattice, which is also puckered. Due to distortion, the squares appear to be corner-connected with alternating small and large squares. The remaining space between two adjacent corner-connected squares forms a distorted rhombus (Figure 2).

*X-ray Photoelectron Spectroscopy:* To determine the oxidation states of the elements in our compound, we collected the XPS data. The survey spectra, as shown in Figure S1(a), indicate the presence of all the constituent elements. By deconvoluting and fitting the core-level spectra for Ru 3d, we identified four peaks (Figure 3a). Two of these peaks, located at approximately 284.6 eV and 289.4 eV, correspond to Ru $3d_{5/2}$ and $3d_{3/2}$, respectively. These binding energy values are similar to those observed for $BaRuO_4$.[36,37] Additionally, the two peaks at around 279 eV and 281 eV are attributed to Sr $2p_{3/2}$ and its satellite peak, respectively.[36–38]

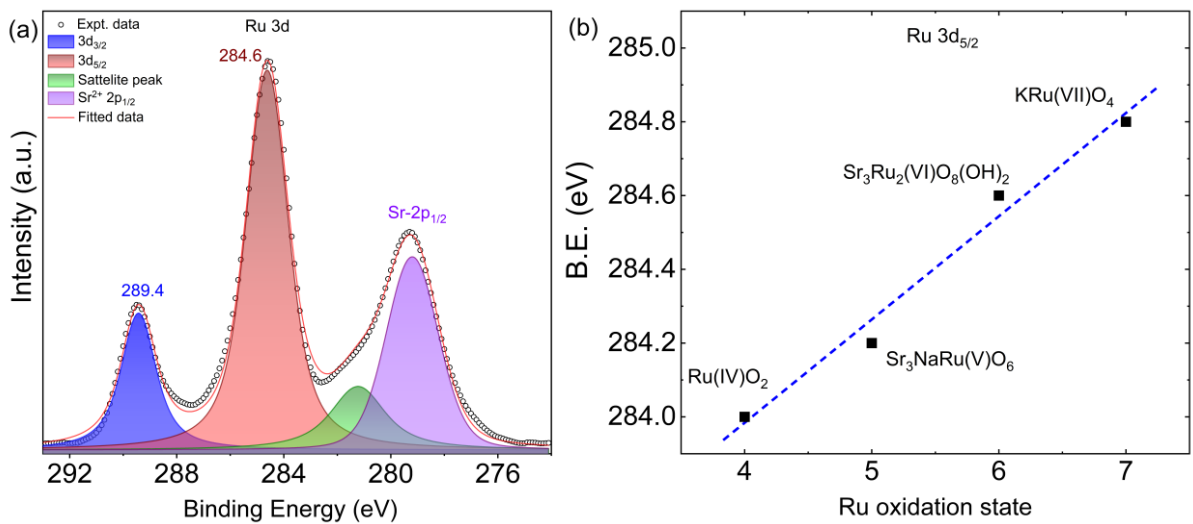

**Figure 3.** (a) Deconvoluted and fitted core-level photoelectron spectra for Ru 3d, (b) Comparison of $3d_{5/2}$ peaks for different Ru oxidation states.

As the Ru 3d spectra are complicated to analyze due to the presence of the C 1s peak and satellite peaks, we have compared them with other Ru oxides. A linear variation of increasing binding energy with increasing oxidation state is obtained which experimentally reaffirms the VI oxidation state of Ru in our compound (Figure 3b). Sr 3d and O 1s spectra of the sample, along with the compared Ru $3d_{5/2}$ spectra are presented in Figure S1.

*Fourier Transformed Infrared Spectroscopy:* To confirm the presence of hydrogen atoms bonded to the axial oxygen atoms in the $RuO_5$ trigonal bipyramids, we recorded the transmittance spectrum on the single crystals, as shown in Figure 4. In a perfect TBP geometry ($D_{3h}$ point group), the number of normal vibrational modes is $3N - 6 = (3 \times 5) - 6 = 9$. These 9 modes are $\Gamma_{vib} = 2A_1' + 2A_2'' + 3E' + 3E''$, out of which $A_2''$ and $E'$ modes are IR active. These bands are generally observed in the low-frequency region. Bands corresponding to the Ru–O and Sr–O bonds are detected below 1000 cm$^{-1}$, making them difficult to identify, as literature on similar structures is limited.[39,40] Furthermore, the O–H bonds demonstrate spectral bands in the range of 3000–3700 cm$^{-1}$ for stretching modes, 1300–1500 cm$^{-1}$ for in-plane bending, and 650–750 cm$^{-1}$ for out-of-plane bending modes.[41] The stretching modes exhibit broad features from hydrogen bonding, while isolated O–H groups show a weak and sharp feature at higher frequencies.[41,42] The reflectance spectrum of our compound shows two weak but sharp bands approximately at 3580 cm$^{-1}$ and 3530 cm$^{-1}$, along with another sharp band around 1350 cm$^{-1}$ strongly suggesting the presence of structural –OH groups.

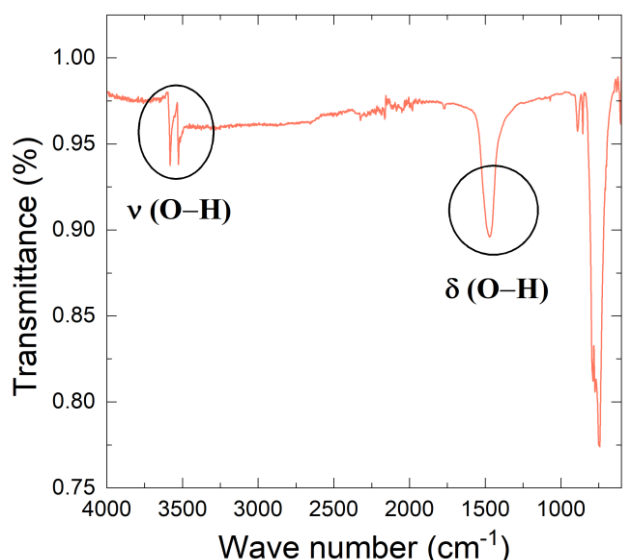

**Figure 4**. FTIR spectrum for $Sr_3Ru_2O_8(OH)_2$

The bands at the higher frequencies are likely related to the stretching of isolated O–H bonds [ν(OH)], while the lower frequency band may be associated with in-plane bending [δ(OH)].[41] The out-of-plane bending mode, which is expected to occur at lower frequencies, overlaps with the lattice modes of the $RuO_5$ trigonal bipyramids, complicating its identification.[39,41] It is important to note that the observed modes exhibit distinctive features, likely due to the influence of the local lattice environment. In the $RuO_5$ trigonal bipyramids, there are five bonds: three equatorial bonds with similar lengths and two axial bonds, one of which is longer than the other. This difference in bond lengths accounts for the presence of two small, weak peaks. Furthermore, the bending modes are primarily influenced by the metal centre associated with the O–H bond, specifically Ru, which explains the sharp band that is observed. A similar characteristic feature is also observed in $Ae_3[Tm(OH)_6]_2$, (where, Ae = Ba, Sr and Ca; Tm = Cr and Rh).[43]

*Thermal Analysis:* The thermal stability of the material was evaluated using a thermogravimetric analyser, which is presented in Figure 5. The samples were heated from room temperature to 800-1000 °C in both nitrogen ($N_2$) and air atmosphere, with the percentage of mass loss recorded. After an initial small weight loss due to adsorbed water, the sample starts decomposing above 650 K, indicating loss of structural water. The total mass loss until 1073K is about 7.5 %. The decomposition product (after heating in a nitrogen atmosphere), identified using PXRD, is mostly $Sr_3Ru_2O_7$ (Figure S2). The decomposition profile can be written as follows: $Sr_3[(RuO_4)OH]_2 \rightarrow Sr_3Ru_2O_7 + H_2O\uparrow + O_2\uparrow$

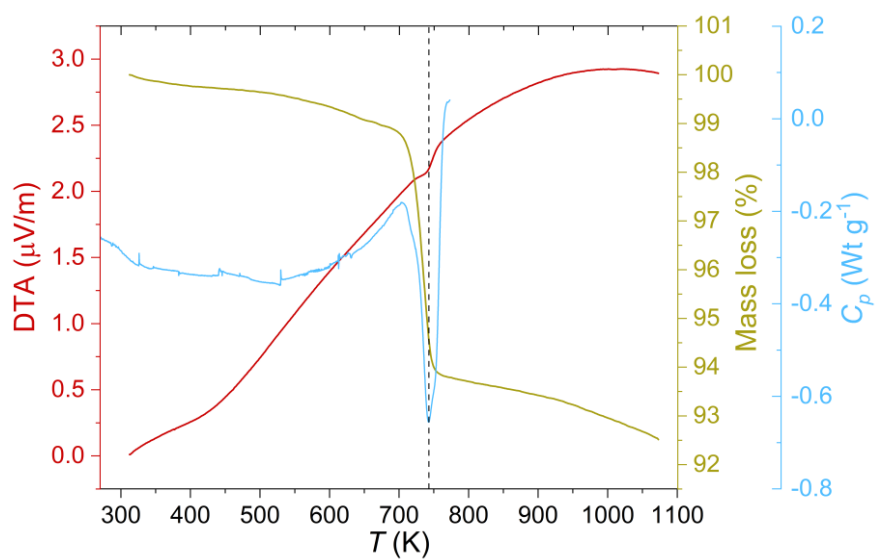

**Figure 5.** Thermal analysis (TGA, DTA and DSC plots) of $Sr_3Ru_2O_8(OH)_2$

The theoretical weight loss following the above equation should be 7.9 %, which matches well with the observed weight loss of 7.5 %. In contrast, the diffractogram recorded after TGA under ambient air atmosphere exhibits broad peaks, which were tentatively assigned to $Sr_3Ru_2O_7$. This absence of notable peaks may indicate the formation of an amorphous product (Figure S2 in the SI).

*Magnetic studies:* Temperature-dependent DC magnetization study of the samples in the presence of an external magnetic field of 5000 Oe is presented in Figure 6. The compound exhibits paramagnetic susceptibility at least until 50 K, below which a plateau is visible. However, there is no discernible magnetic anomaly. At higher temperatures, the compound exhibits Curie-Weiss behaviour. A straight line fit to the inverse molar magnetic susceptibility in the temperature range of 100-300 K yields a Weiss constant $\theta_W \sim -123(1)$ K and a Curie constant of 2.2 emu mol$^{-1}$ K$^{-1}$. A large negative Weiss constant indicates strong antiferromagnetic correlation. The effective paramagnetic moment calculated from the Curie constant is $P_{eff}$ ~2.96 $\mu_B$/Ru$^{6+}$, which is closer to the theoretical magnetic moment calculated using the spin-only formula: $\boldsymbol{\mu_{theo} = \sqrt{[n(n+2)]}}$, where 'n' is the number of unpaired electrons (2.83 $\mu_B$ for n = 2). TBP crystal field leads to the splitting of $d$ orbitals set into three subset; $\mathbf{1e''(xz, yz) < 1e'(x^2 - y^2, xy) < 1a'(z^2)}$, so that the ground state electronic configuration of each Ru$^{6+}$ ($d^2$) site is $d_{xz}^1 d_{yz}^1$ (assuming the $z$-axis is along the Ru−O axial bond). The observed magnetic moment confirms the presence of two unpaired electrons in the $d$-shell, giving rise to a $d^2$ configuration.[44]

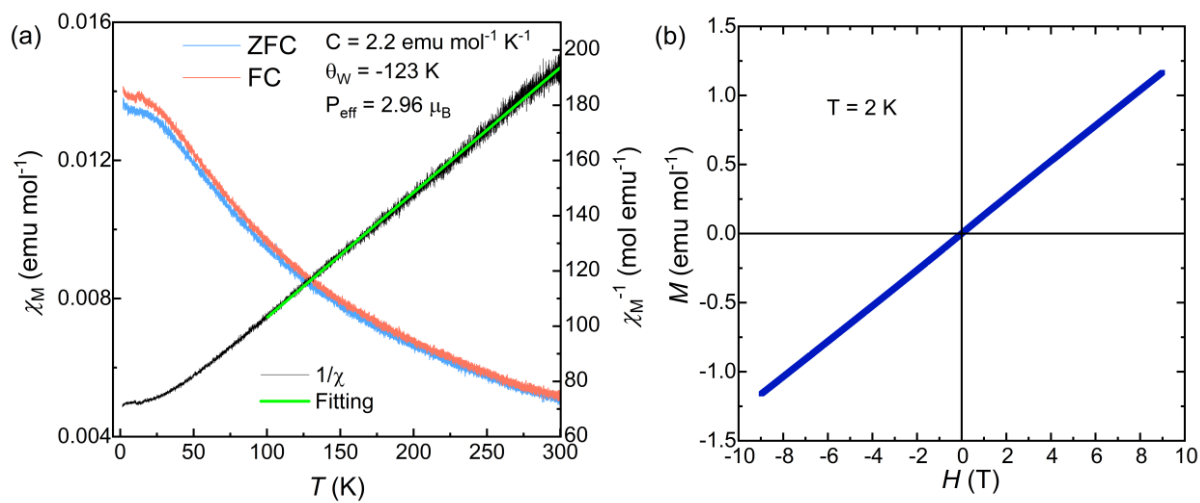

**Figure 6.** (a) Temperature-dependent molar magnetic susceptibility and its inverse (b) Isothermal magnetisation curves recorded at 2 K.

It may be noted that a small cusp-like behaviour is observed at the low temperature region of our susceptibility curve, which may indicate the presence of a magnetic ordering. However, AC susceptibility data did not indicate the presence of any magnetic transition (Figure S3). A linear M-H curve passing through the origin also confirms a paramagnetic state of the compound and the absence of any magnetic ordering.

*Electrochemical Properties:* Electrochemical water splitting comprises two primary half-reactions: oxygen evolution reaction (OER) at the anode and hydrogen evolution reaction (HER) at the cathode. The OER half reaction produces electrons, protons, and oxygen during oxidation process, which are essential components in fuel production processes like $CO_2$ reduction and hydrogen generation.[45] As a result, water oxidation catalysts play a vital role in fuel generation systems. This reaction involves $4e^-$ and $4H^+$ transfer, which causes a high energy barrier and makes the water-splitting process a slow kinetic reaction.[46] $RuO_2$ and $IrO_2$ are two noble metal oxides that demonstrate superior OER catalytic performance but are limited by their high cost, scarcity, and long-term stability.[47] Designing and developing an effective and affordable catalyst for large-scale water splitting is necessary. We have analyzed $Sr_3Ru_2O_8(OH)_2$ for water splitting applications while comparing it to commercial $RuO_2$.

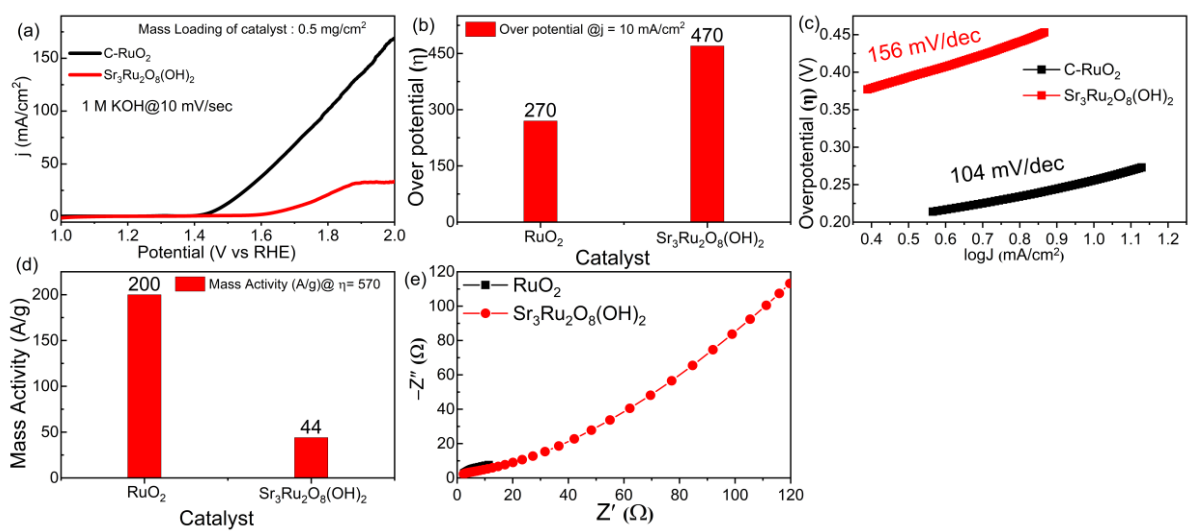

**Figure 7.** Electrocatalytic activity of C-$RuO_2$ and $Sr_3Ru_2O_8(OH)_2$ catalyst materials (a) LSV in 1 M KOH at a scan rate of 10 mV/sec (b) Over potential at j= 10 mA/cm$^2$, (c)Tafel Plot, (d) Mass Activity at η = 570 Mv, (e) Electrochemical Impedance Spectra carried out at a frequency region of 0.01 Hz - 100 KHz in 1 M KOH solution.

Electrochemical OER measurements for $Sr_3Ru_2O_8(OH)_2$ and C-$RuO_2$ were conducted using a CHI electrochemical workstation in a 1 M KOH electrolyte solution. A standard three-electrode configuration was employed, comprising a graphite sheet as the working electrode, a platinum wire as the counter electrode, and an Ag/AgCl electrode (3 M KCl) as the reference electrode. Linear sweep voltammetry (LSV) polarization curves and overpotential data, as presented in Figure 7a and 7b, demonstrate that C-RuO$_2$ exhibits superior oxygen evolution reaction (OER) electrocatalytic performance. It achieves a current density of 10mA/cm$^2$ at an overpotential of 270 mV, which is significantly lower than the Sr$_3$Ru$_2$O$_8$(OH)$_2$ catalyst which required 470 mV. These overpotential values are comparable to those of most other reported OER catalysts.

Tafel plots provide an effective way to evaluate electrocatalyst performance by illustrating the relationship between overpotential and current density. A lower Tafel slope (b), a crucial parameter extracted from these plots, indicates that a smaller overpotential is needed to achieve a specific current density, reflecting faster electron transfer kinetics.[48] The Tafel slopes calculated from the LSV curves, shown in Figure 7c. C-RuO$_2$ had the smaller Tafel slope of 104 mV dec$^{-1}$ compared to the Sr$_3$Ru$_2$O$_8$(OH)$_2$ (156 mV dec$^{-1}$). These values are comparable to those of other reported OER catalysts suggesting that C-RuO$_2$ has an enhanced kinetic pathway, leading to faster electrochemical reaction rates and improved reaction kinetics. To further quantify efficiency, mass activity was calculated at an overpotential of 570 mV. The results, shown in Figure 7d, show that the mass activity of C-RuO$_2$(200 A/g) is nearly four times higher than that of Sr$_3$Ru$_2$O$_8$(OH)$_2$ (44 A/g), confirming its enhanced overall catalytic efficiency.

To further understand how charge transfer influences these kinetics, we performed electrochemical impedance spectroscopy (EIS), with the findings presented in Figure 7e. The results reveal that C-RuO$_2$ catalysts have a notable reduction in charge-transfer resistance (R$_{ct}$) compared to Sr$_3$Ru$_2$O$_8$(OH)$_2$. Since R$_{ct}$ is inversely related to the electron transfer rate, this observed decrease directly correlates with enhanced OER activity, indicating superior electron transfer efficiency and facilitating more efficient electron transfer during the reaction.

The linear sweep voltammetry (LSV) data show that both C-RuO$_2$ and Sr$_3$Ru$_2$O$_8$(OH)$_2$ are active oxygen evolution reaction (OER) electrocatalysts. While C-RuO$_2$ demonstrates superior performance with a lower overpotential of 270 mV to achieve a current density of 10mAcm$^{-2}$, the Sr$_3$Ru$_2$O$_8$(OH)$_2$ catalyst's overpotential of 470 mV is still considered comparable to many other OER catalysts reported in the literature. Furthermore, the above study suggests a conducting nature of our material. In conclusion, an integrated analysis of overpotential, Tafel slopes, and electrochemical impedance spectroscopy (EIS) data confirms that C-RuO$_2$ is more efficient than Sr$_3$Ru$_2$O$_8$(OH)$_2$. However, the performance of Sr$_3$Ru$_2$O$_8$(OH)$_2$ remains relevant and is on par with the broader field of OER catalysts.

*Electronic Band Structure:* The model used for the theoretical calculation was taken from single-crystal data and optimized. In the band structure (Figure 8, left panel), nearly flat bands at the Fermi level indicate a high density of states at that level. This flatness suggests that the electrons associated with these bands have very low group velocity and high effective mass, which may lead to strong electron-electron correlations.

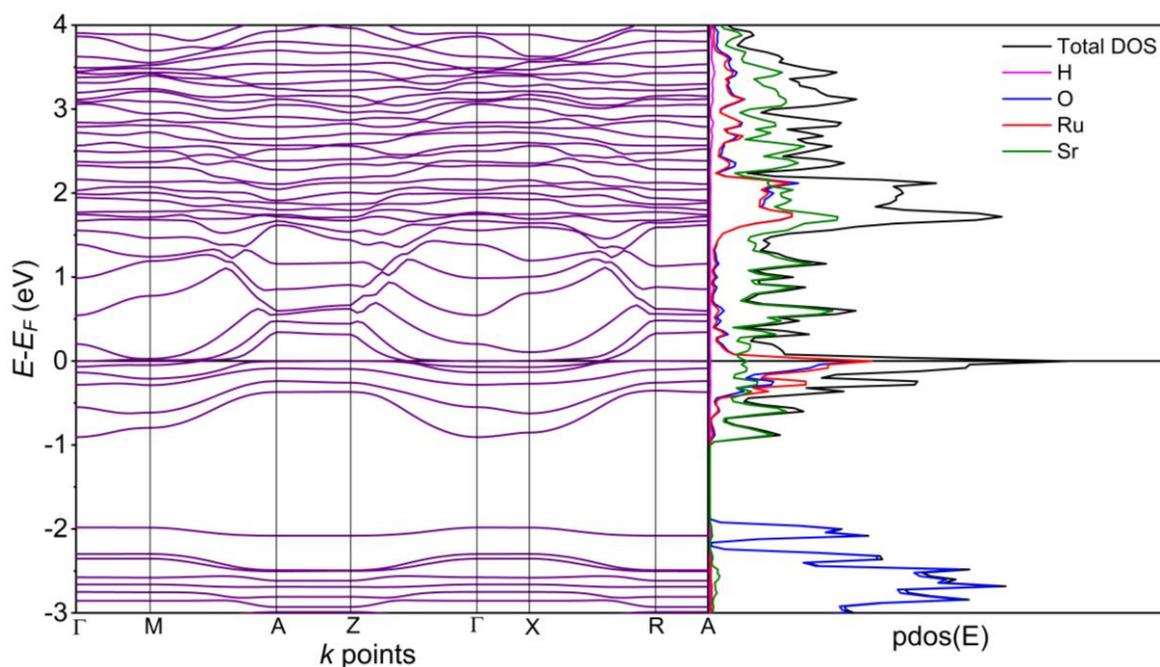

**Figure 8**. Band structure (left panel) and density of electronic states (right panel) of Sr$_3$Ru$_2$O$_8$(OH)$_2$. The DOS plot has partial density of states of Sr, Ru, O, and H with the Fermi level assigned at zero.

In a similar manner, a finite and significant density of states (DOS) at the Fermi level suggests a semi-metallic or metallic characteristic, which aligns with the conducting properties inferred from the electrochemical measurements. In the atom-decomposed density of states (DOS) plot (Figure S4), it is evident that the electronic structure of $Sr_3Ru_2O_8(OH)_2$ is primarily influenced by the hybridization between Ru $d$ and O $p$ orbitals. A similar nature is also observed in the $Ag_2RuO_4$ system.[19] These interactions collectively shape the bands around the Fermi level. Additionally, the well-dispersed states of Sr, Ru, and O suggest that covalent interactions are the dominant factor in this material.

**Conclusions**

A hexavalent ruthenium oxo-hydroxide is synthesized as single crystals and powder under hydrothermal conditions in a strongly basic medium. The structure is solved using single crystal data. The compound crystallizes in a new non-centrosymmetric structure and contains isolated ruthenium trigonal bipyramids. The ruthenium atoms assemble as alternating, puckered, corner-connected distorted squares and edge-sharing rhombus. Combined single crystal, XPS, IR, thermogravimetric and magnetic data confirm a rare hexavalent oxidation state for ruthenium. This is only the second instance where a hexavalent ruthenate is accessed under mild synthesis conditions. The compound is paramagnetic down to 2 K and may be electrically conducting, as evidenced by electrochemical studies and theoretical calculations.

**Note:** During the finalisation of this manuscript, we became aware of another similar report presenting the synthesis, structure, and magnetic properties of compounds reported here.[49] In the publication by Walton et al, a powder sample is obtained, and the structure was determined using *in situ* neutron diffraction and electron diffraction techniques. The structure and characterisation reported there is in principle same as reported here. However, our work highlights the crystal growth and unambiguous structure determination of the $Sr_3Ru_2O_8(OH)_2$ compound using single crystals. We have additionally studied the electrochemical properties

and electronic band structure of Sr$_3$Ru$_2$O$_8$(OH)$_2$, which indicate a metallic nature of the compound.

**Supporting Information.**

The Supplementary material file includes: XPS spectra, PXRD before and after TGA, AC susceptibility plot, pDOS plot, Table of EDX analysis, atomic coordinates, anisotropic thermal parameters and Ru–O bond length refined from SC-XRD data.

**Author Information**

**Corresponding Author**

***Gohil S. Thakur**- Department of Chemical Sciences, Indian Institute of Science Education and Research, Berhampur-760030, Odisha, India; https://orcid.org/0000-0002-1362-2357; Email: gsthakur@iiserbpr.ac.in

**Authors:**

**Subham Naik**- Department of Chemical Sciences, Indian Institute of Science Education and Research, Berhampur-760030, Odisha, India; https://orcid.org/0000-0003-1685-3748

**Soumili Dutta**-Department of Chemical Sciences, Indian Institute of Science Education and Research, Berhampur-760030, Odisha, India; https://orcid.org/0009-0004-5940-9078

**Hiranmayee Senapati**- Department of Chemical Sciences, Indian Institute of Science Education and Research, Berhampur-760030, Odisha, India; https://orcid.org/0009-0006-4978-1292

**Sweta Yadav**- Department of Chemistry, Indian Institute of Technology Hyderabad, Kandi, Sangareddy-502284, Telangana, India; https://orcid.org/0000-0001-7245-2426

**Subarna Ray-** Department of Physical Sciences, Indian Institute of Science Education and Research, Berhampur-760030, Odisha, India;

**Jai Prakash**- Department of Chemistry, Indian Institute of Technology Hyderabad, Kandi, Sangareddy-502284, Telangana, India; https://orcid.org/0000-0002-4078-9662

**Rahul Sharma-**Department of Physical Sciences, Indian Institute of Science Education and Research, Berhampur-760030, Odisha, India;

**Acknowledgement**

GST acknowledges the generous financial support from the institute in the form of seed grant and ANRF for the Early Career Research Grant (sanction no.: ANRF/ECRG/2024/001436/CS). SN thanks IISER Berhampur for a postdoctoral fellowship. The authors from IISERBPR thank the Director, IISER Berhampur, for support during lab establishment, and the Central Advanced Instrumentation Facility (CAIF) for providing infrastructure and instrumental support. We thank Dr. Ravi Kumar Kunchala for his assistance in electrochemical measurements and Mr. Haribrahma Singh for AC susceptibility measurements.

**Conflict of Interest.**

Authors declare no conflicts of interest.

**References.**

(1) Mazin, I. I.; Singh, D. J. Electronic Structure and Magnetism in Ru-Based Perovskites. *Phys. Rev. B,* **1997**, *56* (5), 2556–2571. https://doi.org/10.1103/PhysRevB.56.2556.

(2) Ovchinnikov, S. G. Exotic Superconductivity and Magnetism in Ruthenates. *Phys. Uspekhi,* **2003**, *46* (1), 21–44. https://doi.org/10.1070/PU2003v046n01ABEH001235.

(3) Cao, G.; Alexander, C. S.; McCall, S.; Crow, J. E.; Guertin, R. P. From Antiferromagnetic Insulator to Ferromagnetic Metal: A Brief Review of the Layered Ruthenates. *Mater. Sci. Eng. B,* **1999**, *63* (1–2), 76–82. https://doi.org/10.1016/S0921-5107(99)00055-0.

(4) Bensch, W.; Schmalle, H. W.; Reller, A. Structure and Thermochemical Reactivity of $CaRuO_3$ and $SrRuO_3$. *Solid State Ion.,* **1990**, *43*, 171–177. https://doi.org/10.1016/0167-2738(90)90481-6.

(5) Müller-Buschbaum, Hk.; Wilkens, J. Ein Beitrag Über $Sr_2RuO_4$ Und $Sr_3Ru_2O_7$ Zur Oktaederstreckung von $M^{4+}$ in $K_2NiF_4$ - Und $Sr_3Ti_2O_7$ -Typ-Verbindungen. *Z. Anorg. Allg. Chem.,* **1990**, *591* (1), 161–166. https://doi.org/10.1002/zaac.19905910118.

(6) Hiley, C. I.; Lees, M. R.; Fisher, J. M.; Thompsett, D.; Agrestini, S.; Smith, R. I.; Walton, R. I. Ruthenium(V) Oxides from Low-Temperature Hydrothermal Synthesis. *Angew. Chem. Int. Ed.,* **2014**, *53* (17), 4423–4427. https://doi.org/10.1002/anie.201310110.

(7) Marchandier, T.; Jacquet, Q.; Rousse, G.; Baptiste, B.; Abakumov, A. M.; Tarascon, J.-M. Expanding the Rich Crystal Chemistry of Ruthenium(V) Oxides via the Discovery of $BaRu_2O_6$, $Ba_5Ru_4O_{15}$, $Ba_2Ru_3O_{10}$, and $Sr_2Ru_3O_9(OH)$ by PH-Controlled Hydrothermal Synthesis. *Chem. Mater.,* **2019**, *31* (16), 6295–6305. https://doi.org/10.1021/acs.chemmater.9b02510.

(8)  Prasad, B. E.; Sadhukhan, S.; Hansen, T. C.; Felser, C.; Kanungo, S.; Jansen, M. Synthesis, Crystal and Magnetic Structure of the Spin-Chain Compound $Ag_2RuO_4$. *Phys. Rev. Mater.* **2020**, *4* (2), 024418. https://doi.org/10.1103/PhysRevMaterials.4.024418.

(9)  Prasad, B. E.; Kanungo, S.; Jansen, M.; Komarek, A. C.; Yan, B.; Manuel, P.; Felser, C. $AgRuO_3$, a Strongly Exchange-Coupled Honeycomb Compound Lacking Long-Range Magnetic Order. *Chem. Eur. J.,* **2017**, *23* (19), 4680–4686. https://doi.org/10.1002/chem.201606057.

(10) Prasad, B. E.; Kazin, P.; Komarek, A. C.; Felser, C.; Jansen, M. B-$Ag_3RuO_4$, a Ruthenate(V) Featuring Spin Tetramers on a Two-Dimensional Trigonal Lattice. *Angew. Chem.,* **2016**, *128* (14), 4543–4547. https://doi.org/10.1002/ange.201510576.

(11) Cordfunke, E. H. P.; Van Der Laan, R. R.; Westrum, E. F. The Thermochemical and Thermophysical Properties of $Cs_2RuO_4$ and $Cs_2MnO_4$ at Temperatures from 5 K to 1000 K. *J. Chem. Thermodyn.* **1992**, *24* (8), 815–822. https://doi.org/10.1016/S0021-9614(05)80226-5.

(12) Fischer, D.; Hoppe, R.; Mogare, K. M.; Jansen, M. Syntheses, Crystal Structures and Magnetic Properties of $Rb_2RuO_4$ and $K_2RuO_4$. *Z. Naturforsch. B,* **2005**, *60* (11), 1113–1117. https://doi.org/10.1515/znb-2005-1101.

(13) Mogare, K. M.; Klein, W.; Peters, E. M.; Jansen, M. $K_3Na(RuO_4)_2$ and $Rb_3Na(RuO_4)_2$, Two New Ruthenates with Glaserite Structure. *Solid State Sci.,* **2006**, *8* (5), 500–507. https://doi.org/10.1016/J.SOLIDSTATESCIENCES.2006.01.004.

(14) Mogare, K. M.; Friese, K.; Klein, W.; Jansen, M. Syntheses and Crystal Structures of Two Sodium Ruthenates: $Na_2RuO_4$ and $Na_2RuO_3$. *Z. Anorg. Allg. Chem.,* **2004**, *630* (4), 547–552. https://doi.org/10.1002/zaac.200400012.

(15) Nowogrocki, G.; Abraham, F.; Tréhoux, J.; Thomas, D. Configuration de l'ion Ruthénate: Structure Cristalline Du Dihydroxotrioxoruthénate(VI) de Baryum, $Ba[RuO_3(OH)_2]$. *Acta Crystallogr. B,* **1976**, *32* (8), 2413–2419. https://doi.org/10.1107/S0567740876007875.

(16) Fischer, D.; Hoppe, R. Zur Konstitution von Alkaliruthenaten (VI). 2. Über Den Aufbau von $K_2[RuO_3(OH_2]$. *Z. Anorg. Allg. Chem.,* **1991**, *601* (1), 41–46. https://doi.org/10.1002/zaac.19916010105.

(17) Hansen, T.; Le Bail, A.; Laligant, Y. Synthesis and Structure Approach of Barium-Oxomercurato(II)-Oxoruthenate(VI) $BaHgRuO_5$. *J. Solid State Chem.,* **1995**, *120* (2), 223–230. https://doi.org/10.1006/jssc.1995.1402.

(18) Fischer, D.; Hoppe, R. Über „Gemischt-Koordinierte" Einkernige Anionen. 2. Ein Oxoruthenat(VI) Neuen Typs: $CsK_5Ru_2O_9=CsK_5[RuO_5][RuO_4]$. *Z. Anorg. Allg. Chem.,* **1992**, *617* (11), 37–44. https://doi.org/10.1002/zaac.19926170107.

(19) Prasad, B. E.; Sadhukhan, S.; Hansen, T. C.; Felser, C.; Kanungo, S.; Jansen, M. Synthesis, Crystal and Magnetic Structure of the Spin-Chain Compound $Ag_2RuO_4$. *Phys. Rev. Mater.,* **2020**, *4* (2), 024418. https://doi.org/10.1103/PhysRevMaterials.4.024418.

(20) Bruker. APEX 3. Bruker AXS Inc.: Madison, Wisconsin, USA. **2012**.

(21) Sheldrick, G. M. *SHELXT* – Integrated Space-Group and Crystal-Structure Determination. *Acta Crystallogr. A Found. Adv.,* **2015**, *71* (1), 3–8. https://doi.org/10.1107/S2053273314026370.

(22) Sheldrick, G. M. Crystal Structure Refinement with *SHELXL*. *Acta Crystallogr. C Struct. Chem.* **2015**, *71* (1), 3–8. https://doi.org/10.1107/S2053229614024218.

(23) Sheldrick, G. M.; Schneider, T. R. SHELXL: High-Resolution Refinement; **1997**; 319–343. https://doi.org/10.1016/S0076-6879(97)77018-6.

(24) Sheldrick, G. M. SADABS, Program for Area Detector Adsorption Correction. Institute for Inorganic Chemistry, University of Gottingen, Germany, **1996**.

(25) Moulder, J. F.; Stickle, W. F.; Sobol, P. E.; Bomben, K. D. *Handbook of X-Ray Photoelectron Spectroscopy: A Reference Book of Standard Spectra for Identification and Interpretation of XPS Data*, 5th ed.; Chastain Jill, Ed.; Perkin-Elmer Corporation, Physical Electronics Division: Minnesota, USA, **1992**.

(26) Giannozzi, P.; Baseggio, O.; Bonfà, P.; Brunato, D.; Car, R.; Carnimeo, I.; Cavazzoni, C.; de Gironcoli, S.; Delugas, P.; Ferrari Ruffino, F.; Ferretti, A.; Marzari, N.; Timrov, I.; Urru, A.; Baroni, S. Quantum ESPRESSO toward the Exascale. *J. Chem. Phys.,* **2020**, *152* (15). https://doi.org/10.1063/5.0005082.

(27) Giannozzi, P.; Baroni, S.; Bonini, N.; Calandra, M.; Car, R.; Cavazzoni, C.; Ceresoli, D.; Chiarotti, G. L.; Cococcioni, M.; Dabo, I.; Dal Corso, A.; de Gironcoli, S.; Fabris, S.; Fratesi, G.; Gebauer, R.; Gerstmann, U.; Gougoussis, C.; Kokalj, A.; Lazzeri, M.; Martin-Samos, L.; Marzari, N.; Mauri, F.; Mazzarello, R.; Paolini, S.; Pasquarello, A.; Paulatto, L.; Sbraccia, C.; Scandolo, S.; Sclauzero, G.; Seitsonen, A. P.; Smogunov, A.; Umari, P.; Wentzcovitch, R. M. QUANTUM ESPRESSO: A Modular and Open-Source Software Project for Quantum Simulations of Materials. *J. Phys. Condens. Matter.,* **2009**, *21* (39), 395502. https://doi.org/10.1088/0953-8984/21/39/395502.

(28) Giannozzi, P.; Andreussi, O.; Brumme, T.; Bunau, O.; Buongiorno Nardelli, M.; Calandra, M.; Car, R.; Cavazzoni, C.; Ceresoli, D.; Cococcioni, M.; Colonna, N.; Carnimeo, I.; Dal Corso, A.; de Gironcoli, S.; Delugas, P.; DiStasio, R. A.; Ferretti, A.; Floris, A.; Fratesi, G.; Fugallo, G.; Gebauer, R.; Gerstmann, U.; Giustino, F.; Gorni, T.; Jia, J.; Kawamura, M.; Ko, H.-Y.; Kokalj, A.; Küçükbenli, E.; Lazzeri, M.; Marsili, M.; Marzari, N.; Mauri, F.; Nguyen, N. L.; Nguyen, H.-V.; Otero-de-la-Roza, A.; Paulatto, L.; Poncé, S.; Rocca, D.; Sabatini, R.; Santra, B.; Schlipf, M.; Seitsonen, A. P.; Smogunov, A.; Timrov, I.; Thonhauser, T.; Umari, P.; Vast, N.; Wu, X.; Baroni, S. Advanced Capabilities for Materials Modelling with Quantum ESPRESSO. *J. Phys. Condens. Matter.* **2017**, *29* (46), 465901. https://doi.org/10.1088/1361-648X/aa8f79.

(29) Perdew, J. P.; Burke, K.; Ernzerhof, M. Generalized Gradient Approximation Made Simple. *Phys. Rev. Lett.,* **1996**, *77* (18), 3865–3868. https://doi.org/10.1103/PhysRevLett.77.3865.

(30) Monkhorst, H. J.; Pack, J. D. Special Points for Brillouin-Zone Integrations. *Phys. Rev. B,* **1976**, *13* (12), 5188–5192. https://doi.org/10.1103/PhysRevB.13.5188.

(31) Momma, K.; Izumi, F. *VESTA*: A Three-Dimensional Visualization System for Electronic and Structural Analysis. *J. Appl. Crystallogr.,* **2008**, *41* (3), 653–658. https://doi.org/10.1107/S0021889808012016.

(32) Popova, T. L.; Kisel, N. G.; Karlov, V. P.; Krivobok, V. I. Bivalent Metal Ruthenates (VI). *Russ. J. Inorg. Chem. (Engl. Trans.),* **1981**, *26*, 1613–1615.

(33) Nowogrocki, G.; Abraham, F.; Tréhoux, J.; Thomas, D. Configuration de l'ion Ruthénate: Structure Cristalline Du Dihydroxotrioxoruthénate(VI) de Baryum, Ba[RuO$_3$(OH)$_2$]. *Acta Crystallogr. B,* **1976**, *32* (8), 2413–2419. https://doi.org/10.1107/S0567740876007875.

(34) Hansen, T. W.; Le Bail, A.; Laligant, Y. Barium-Oxomercurato(II)-Oxoruthenate(VI) BaHgRuO$_5$: A New Oxomercurate with a Cyclic Mercurate-Ruthenate Anion High Pressure

Synthesis and Ab Initio Structure Approach by X-Ray Powder Diffraction. *Mater. Sci. Forum,* **1996**, *228–231*, 729–734. https://doi.org/10.4028/www.scientific.net/MSF.228-231.729.

(35) Mogare, K. M.; Friese, K.; Klein, W.; Jansen, M. Syntheses and Crystal Structures of Two Sodium Ruthenates: $Na_2RuO_4$ and $Na_2RuO_3$. *Z. Anorg. Allg. Chem.,* **2004**, *630* (4), 547–552. https://doi.org/10.1002/zaac.200400012.

(36) Ohyoshi, A.; Gotzfried, F.; Beck, W. POLYNUCLEAR CARBONYL COMPLEXES OF RUTHENIUM AND OSMIUM WITH METHYLTHIOLATE AND BROMIDE BRIDGING LIGANDS. *Chem. Lett.,* **1980**, 1537–1540.

(37) Augustynski, J.; Balsenc, L.; Hinden, J. X-Ray Photoelectron Spectroscopic Studies of $RuO_2$-Based Film Electrodes. *J. Electrochem. Soc.,* **1978**, *125* (7), 1093–1097.

(38) Lee, A. Y.; Powell, C. J.; Gorham, J. M.; Morey, A.; Scott, J. H. J.; Hanisch, R. J. Development of the NIST X-Ray Photoelectron Spectroscopy (XPS) Database, Version 5. *Data. Sci. J.,* **2024**, *23*, 45. https://doi.org/10.5334/dsj-2024-045.

(39) Hameka, H. F.; Jensen, J. O.; Kay, J. G.; Rosenthal, C. M.; Zimmerman, G. L. Theoretical Prediction of Geometries and Vibrational Infrared of Ruthenium Oxide Molecules. *J. Mol. Spectrosc.,* **1991**, *150* (1), 218–221. https://doi.org/10.1016/0022-2852(91)90204-N.

(40) Kazuo Nakamoto. Applications in Inorganic Chemistry. In *Infrared and Raman Spectra of Inorganic and Coordination Compounds : Theory and Applications in Inorganic Chemistry*; Wiley: Hoboken, New Jersey, **2008**; Part A, pp 149–354. https://doi.org/10.1002/9780470405840.ch2.

(41) Hadjiivanov, K. Identification and Characterization of Surface Hydroxyl Groups by Infrared Spectroscopy. In *Advances in Catalysis*; Academic Press Inc., **2014**; Vol. 57, pp 99–318. https://doi.org/10.1016/B978-0-12-800127-1.00002-3.

(42) Dai, F.; Zhuang, Q.; Huang, G.; Deng, H.; Zhang, X. Infrared Spectrum Characteristics and Quantification of OH Groups in Coal. *ACS Omega,* **2023**, *8* (19), 17064–17076. https://doi.org/10.1021/acsomega.3c01336.

(43) Albrecht, R.; Bretschneider, V.; Doert, T.; Ruck, M. $Ba_3[Rh(OH)_6]_2 \cdot H_2O$–a Precursor to Barium Oxorhodates with One-Dimensional Hydrogen Bonding Systems. *Z. Anorg. Allg. Chem.* **2021**, *647* (18), 1702–1708. https://doi.org/https://doi.org/10.1002/zaac.202100016.

(44) Amoretti, G.; Fournier, J. M. On the Interpretation of Magnetic Susceptibility Data by Means of a Modified Curie-Weiss Law. *J. Magn. Magn. Mater.,* **1984**, *43* (3), L217–L220. https://doi.org/10.1016/0304-8853(84)90069-6.

(45) Wang, W.; Xu, M.; Xu, X.; Zhou, W.; Shao, Z. Perovskite Oxide Based Electrodes for High-Performance Photoelectrochemical Water Splitting. *Angew. Chem. Int. Ed.,* **2020**, *59* (1), 136–152. https://doi.org/10.1002/anie.201900292.

(46) Sun, H.; Zhu, Y.; Jung, W. Tuning Reconstruction Level of Precatalysts to Design Advanced Oxygen Evolution Electrocatalysts. *Molecules,* **2021**, *26* (18), 42–45. https://doi.org/10.3390/molecules26185476.

(47) Pascuzzi, M. E. C.; Man, A. J. W.; Goryachev, A.; Hofmann, J. P.; Hensen, E. J. M. Investigation of the Stability of NiFe-(Oxy) Hydroxide Anodes in Alkaline Water Electrolysis under Industrially Relevant Conditions. *Catal. Sci. Technol.,* **2020**, *10* (16), 5593–5601.

(48) Petrii, O. A.; Tsirlina, G. A. Electrocatalytic Activity Prediction for Hydrogen Electrode Reaction: Intuition, Art, Science. *Electrochim. Acta,* **1994**, *39* (11–12), 1739–1747.

(49) Crossman, M.; Hiley, C. I.; Playford, H. Y.; Smith, R. I.; Hansen, T. C.; Tidey, J. P.; Walton, R. I. Aqueous Synthesis of Strontium Ruthenate(VI) Oxyhydroxides and Their Crystal Structure Solution from Microcrystals. *Inorg. Chem.,* **2025**, *64* (36), 18471–18478.

Supplementary Information

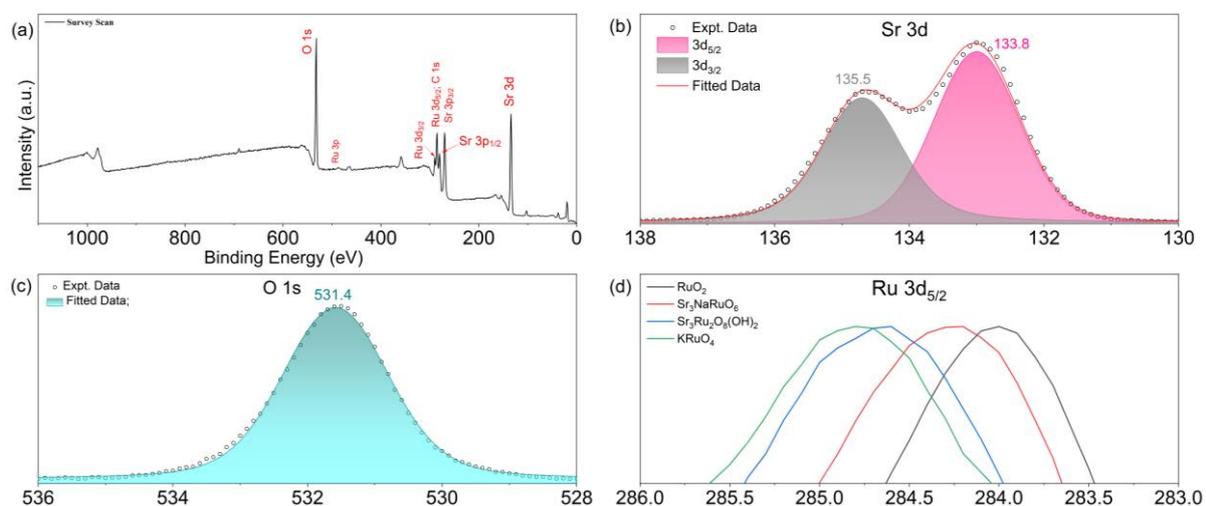

**Figure S1**: (a) XPS survey spectra; (b) fitted core-level Sr 3d spectra; (c) fitted core-level O 1s spectra of the compound; (d) Ru $3d_{5/2}$ spectra for different Ru oxidation states. It should be noted that the XPS spectra of the other Ru compounds were recorded in the same instrument using commercial-grade materials ($RuO_2$ and $KRuO_4$).

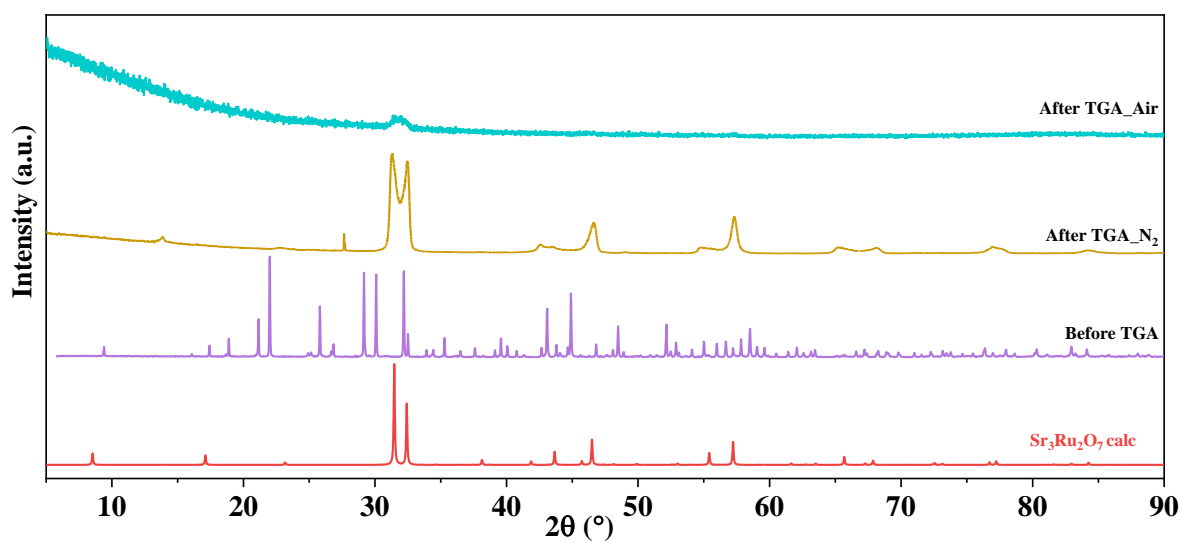

**Figure S2**: PXRD patterns of the compound before and after TGA in comparison to the calculated pattern of $Sr_3Ru_2O_7$.[3,4]

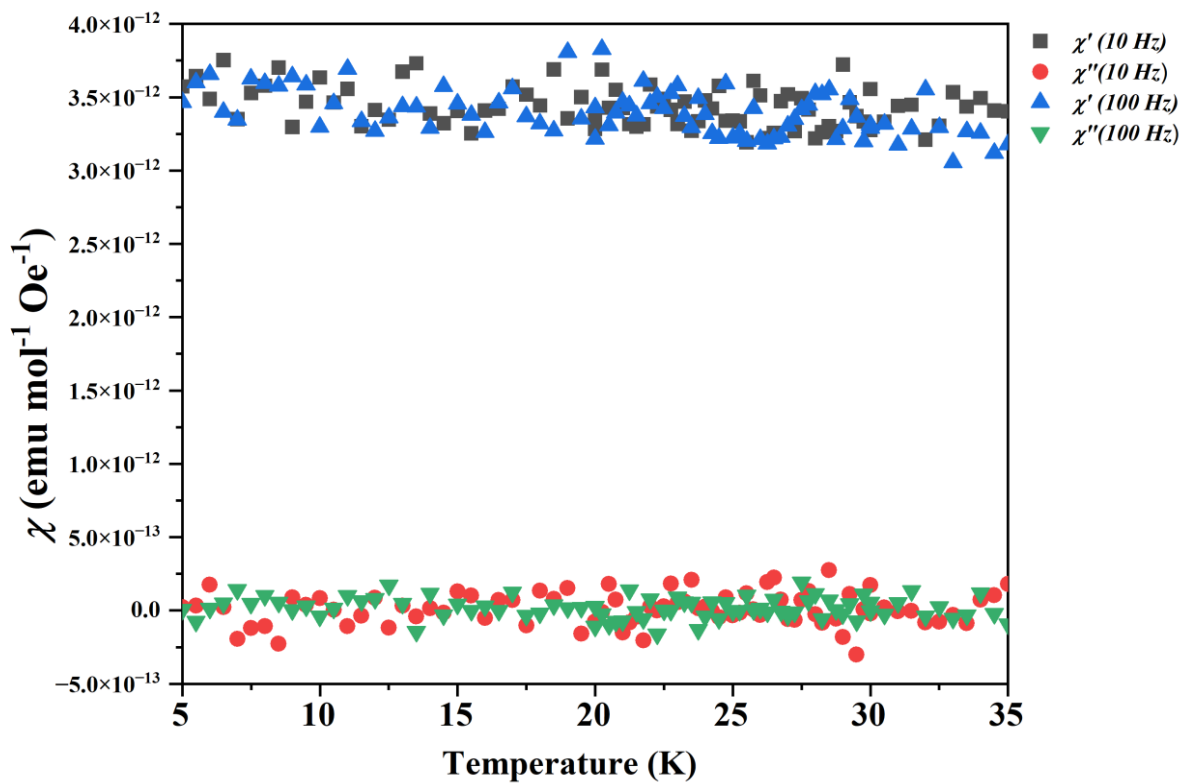

**Figure S3**: AC Susceptibility plots as a function of temperature in the low temperature region at 10 Hz and 100 Hz frequency, showing the absence of any transition.

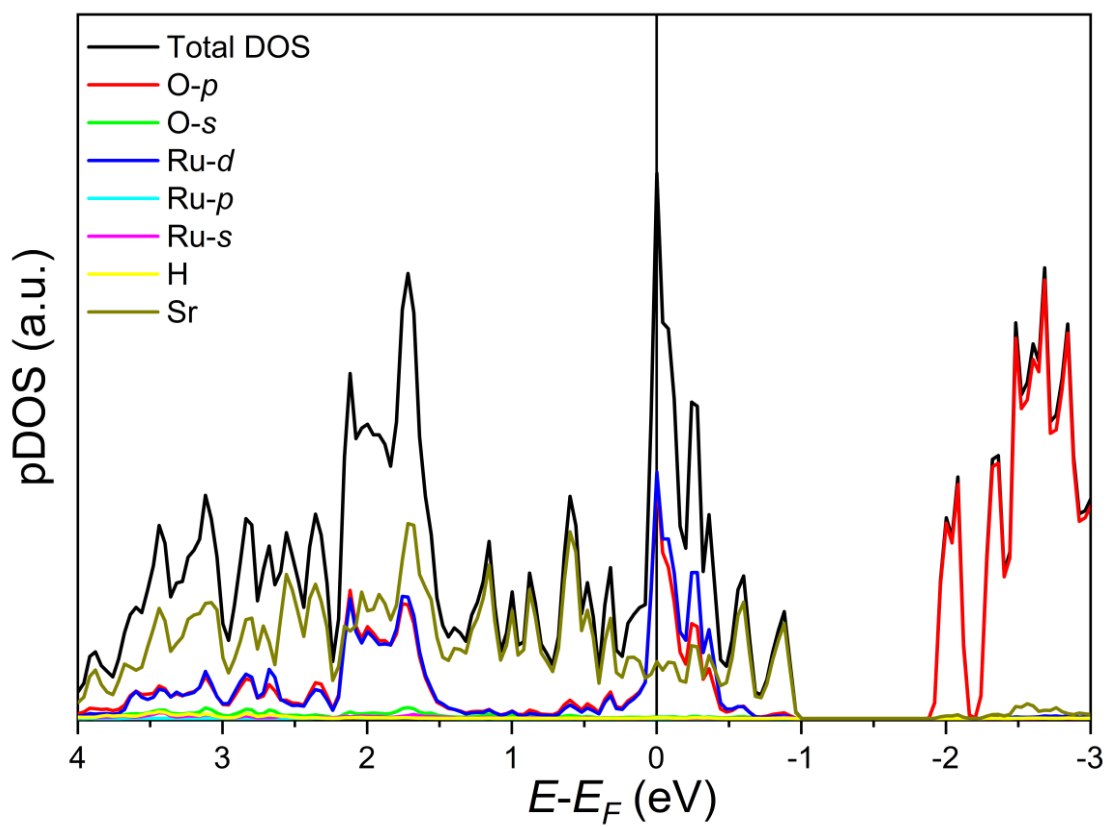

**Figure S4**: Total DOS and atom decomposed pDOS of Ru, O, Sr and H.

## Tables

**Table S1**: Elemental composition obtained from EADX spectra of a single crystal.

| Element | Elemental % at different points | | | | | | | Average Ratio |
|---|---|---|---|---|---|---|---|---|
|  | 1 | 2 | 3 | 4 | 5 | 6 | 7 |  |
| Sr | 15.64 | 15.52 | 14.25 | 15.83 | 19.1 | 19.69 | 17.56 |  |
| Ru | 11.49 | 11.13 | 10.23 | 11.39 | 13.77 | 14.27 | 12.76 | Sr:Ru ~ 1.5 |
| O | 72.87 | 73.35 | 75.52 | 72.78 | 67.13 | 66.04 | 69.68 |  |

**Table S2**: Atomic coordinates and isotropic thermal parameters for $Sr_3Ru_2O_8(OH)_2$.

| Atom | Wyck. | x | y | z | $U_{iso}$ (*$10^{-2}$) (Å$^2$) |
|---|---|---|---|---|---|
| Sr1 | 1b | 0 | 0 | −1/2 | 1.04(2) |
| Sr2 | 4h | −0.19566(4) | 0.10985(4) | 0.00557(11) | 0.907(13) |
| Sr3 | 2g | −1/2 | 0 | 0.04525(18) | 1.49(2) |
| Sr4 | 4h | −0.59754(4) | 0.29224(4) | −0.50509(11) | 0.904(13) |
| Sr5 | 1c | −1/2 | 1/2 | 0 | 1.02(2) |
| Ru1 | 4h | −0.41637(4) | 0.25525(4) | −0.06145(9) | 0.653(11) |
| Ru2 | 4h | −0.25623(4) | −0.07034(4) | −0.43712(9) | 0.777(11) |
| O1 | 4h | −0.3772(4) | −0.1108(4) | −0.6760(10) | 1.91(12) |
| O2 | 4h | −0.2593(4) | −0.1936(4) | −0.3154(10) | 1.66(12) |
| O3 | 4h | −0.1849(4) | −0.0465(4) | −0.7104(9) | 1.42(11) |
| O4 | 4h | −0.1441(4) | −0.0366(4) | −0.2364(8) | 0.91(10) |
| O5 | 4h | −0.3364(4) | 0.0187(4) | −0.2994(10) | 2.16(12) |
| O6 | 4h | −0.2932(4) | 0.2518(4) | −0.1888(10) | 1.64(11) |
| O7 | 4h | −0.3741(4) | 0.1336(4) | 0.1652(9) | 1.17(10) |
| O8 | 4h | −0.5110(4) | 0.1749(4) | −0.1782(10) | 1.57(11) |
| O9 | 4h | −0.4534(4) | 0.3625(4) | −0.2690(9) | 0.98(11) |
| O10 | 4h | −0.4394(4) | 0.3256(4) | 0.2114(9) | 1.14(10) |
| H1 | 4h | −0.352420 | −0.122480 | −0.837160 | 0.023* |
| H7 | 4h | −0.347900 | 0.151590 | 0.322900 | 0.014* |

**Table S3**: Anisotropic thermal parameters for $Sr_3Ru_2O_8(OH)_2$.

| Atom | $U_{11}$ | $U_{22}$ | $U_{33}$ | $U_{12}$ | $U_{13}$ | $U_{23}$ |
|---|---|---|---|---|---|---|
| Sr1 | 0.0105 (3) | 0.0105 (3) | 0.0101 (6) | 0.000 | 0.000 | 0.000 |
| Sr2 | 0.0091 (3) | 0.0092 (3) | 0.0089 (3) | 0.0016 (2) | −0.0011 (2) | −0.0012 (2) |
| Sr3 | 0.0130 (4) | 0.0079 (4) | 0.0238 (5) | −0.0021 (3) | 0.000 | 0.000 |
| Sr4 | 0.0089 (3) | 0.0096 (3) | 0.0086 (3) | −0.0011 (2) | −0.0017 (2) | 0.0015 (2) |
| Sr5 | 0.0098 (3) | 0.0098 (3) | 0.0109 (6) | 0.000 | 0.000 | 0.000 |
| Ru1 | 0.0065 (2) | 0.0064 (2) | 0.0066 (2) | −0.00017 (18) | −0.00001 (19) | 0.00026 (19) |
| Ru2 | 0.0086 (2) | 0.0071 (2) | 0.0076 (2) | −0.00006 (18) | −0.0005 (2) | 0.0004 (2) |
| O1 | 0.017 (3) | 0.017 (3) | 0.023 (3) | −0.006 (2) | −0.012 (2) | 0.005 (2) |
| O2 | 0.024 (3) | 0.010 (3) | 0.016 (3) | −0.005 (2) | −0.008 (2) | 0.006 (2) |
| O3 | 0.018 (3) | 0.017 (3) | 0.008 (2) | 0.000 (2) | 0.001 (2) | 0.0026 (19) |
| O4 | 0.008 (2) | 0.010 (3) | 0.009 (3) | −0.0036 (19) | −0.0035 (18) | 0.0014 (18) |
| O5 | 0.017 (3) | 0.022 (3) | 0.026 (3) | 0.008 (2) | 0.000 (2) | −0.007 (2) |
| O6 | 0.014 (3) | 0.020 (3) | 0.016 (3) | 0.006 (2) | 0.006 (2) | 0.010 (2) |
| O7 | 0.012 (2) | 0.010 (2) | 0.013 (2) | 0.0021 (19) | −0.001 (2) | 0.0021 (19) |
| O8 | 0.018 (3) | 0.012 (3) | 0.017 (3) | −0.005 (2) | −0.007 (2) | 0.005 (2) |
| O9 | 0.008 (2) | 0.009 (3) | 0.013 (3) | 0.0012 (19) | −0.0016 (18) | 0.0061 (19) |
| O10 | 0.016 (3) | 0.012 (3) | 0.007 (2) | 0.000 (2) | 0.0031 (18) | −0.0019 (19) |

**Table S4.** Ru–O Bond lengths in $Sr_3[RuO_4(OH)]_2$.

| Bond | Distance (Å) | Bond | Distance (Å) |
|---|---|---|---|
| Ru1–O6 | 1.776(5) x 1 | Ru1–O1 | 2.140(5) x 1 |
| Ru1–O7 | 2.113(5) x 1 | Ru1–O2 | 1.765(5) x 1 |
| Ru1–O8 | 1.765(5) x 1 | Ru1–O3 | 1.802(5) x 1 |
| Ru1–O9 | 1.887(5) x 1 | Ru1–O4 | 1.904(5) x 1 |
| Ru1–O10 | 1.792(5) x 1 | Ru1–O5 | 1.759(5) x 1 |

**References:**

1   G. M. Sheldrick, *Acta Crystallogr C Struct Chem*, 2015, **71**, 3–8.

2   G. M. Sheldrick and T. R. Schneider, 1997, pp. 319–343.

3   Hk. Müller-Buschbaum and J. Wilkens, *Z Anorg Allg Chem*, 1990, **591**, 161–166.

4   W. Bensch, H. W. Schmalle and A. Reller, *Solid State Ion*, 1990, **43**, 171–177.